\documentclass[sigconf, nonacm, screen]{acmart}
\AtBeginDocument{%
  }

\usepackage{xcolor}
\definecolor{colour3}{RGB}{178,55,250} 

\usepackage{subcaption}
\usepackage{amsmath}
\usepackage{makecell} 
\usepackage{array}  
\usepackage{booktabs}
\usepackage{longtable}
\usepackage{tabularx}
\usepackage{lipsum}

\begin{document}

\title{JaxMARL-HFT: GPU-Accelerated Large-Scale Multi-Agent Reinforcement Learning for High-Frequency Trading}

\author{Valentin Mohl}
\authornote{Authors contributed equally to this research.}
\email{valentin.mohl@cs.ox.ac.uk}
\affiliation{%
  \institution{Department of Computer Science, University of Oxford}
  \country{United Kingdom}
}

\author{Sascha Frey}
\authornotemark[1]
\affiliation{%
  \institution{Department of Computer Science, University of Oxford}
  \country{United Kingdom}}

\author{Reuben Leyland}
\authornotemark[1]
\affiliation{%
  \institution{Department of Engineering Science, University of Oxford}
  \country{United Kingdom}
}

\author{Kang Li}
\affiliation{%
  \institution{Department of Statistics, University of Oxford}
  \country{United Kingdom}
}

\author{George Nigmatulin}
\affiliation{%
 \institution{Department of Engineering Science, University of Oxford}
 \country{United Kingdom}}
 
\author{Mihai Cucuringu}
\affiliation{%
\institution{Department of Mathematics, University of California Los Angeles}
\country{United States} \\ 
\institution{Department of Statistics \& Oxford Man Institute of Quantitative Finance, University of Oxford}
\country{United Kingdom}
}

\author{Stefan Zohren}
\affiliation{%
 \institution{Department of Engineering Science, University of Oxford}
 \country{United Kingdom}}

\author{Jakob Foerster}
\affiliation{%
 \institution{Foerster Lab for AI Research, University of Oxford}
 \country{United Kingdom}}

\author{Anisoara Calinescu}
\affiliation{%
 \institution{Department of Computer Science, University of Oxford}
 \country{United Kingdom}}

\renewcommand{\shortauthors}{Mohl, Frey, Leyland et al.}

\begin{abstract}

Agent-based modelling (ABM) approaches for high-frequency financial markets are difficult to calibrate and validate, partly due to the large parameter space created by defining fixed agent policies. Multi-agent reinforcement learning (MARL) enables more realistic agent behaviour and reduces the number of free parameters, but the heavy computational cost has so far limited research efforts. To address this, we introduce JaxMARL-HFT (\textit{\textbf{JAX}-based \textbf{M}ulti-\textbf{A}gent \textbf{R}einforcement \textbf{L}earning for \textbf{H}igh-\textbf{F}requency \textbf{T}rading}),
 the first GPU-accelerated open-source multi-agent reinforcement learning environment for high-frequency trading (HFT) on market-by-order (MBO) data. Extending the JaxMARL framework and building on the JAX-LOB implementation, JaxMARL-HFT is designed to handle a heterogeneous set of agents, enabling diverse observation/action spaces and reward functions. It is designed flexibly, so it can also be used for single-agent RL, or extended to act as an ABM with fixed-policy agents. Leveraging JAX enables up to a 240x reduction in end-to-end training time, compared with state-of-the-art reference implementations on the same hardware. This significant speed-up makes it feasible to exploit the large, granular datasets available in high-frequency trading, and to perform the extensive hyperparameter sweeps required for robust and efficient MARL research in trading. We demonstrate the use of JaxMARL-HFT with independent Proximal Policy Optimization (IPPO) for a two-player environment, with an order execution and a market making agent, using one year of LOB data (400 million orders), and show that these agents learn to outperform standard benchmarks. The code for the JaxMARL-HFT framework is available on \href{https://github.com/vmohl/JaxMARL-HFT}{GitHub}\footnote{\url{https://github.com/vmohl/JaxMARL-HFT}}.

\end{abstract}




\maketitle

\section{Introduction}

Modelling financial markets at high frequency is an inherently difficult problem, as prices emerge endogenously from the interactions of millions of market participants, often acting with bounded rationality and imperfect information. Agent-based modelling (ABM) addresses this by specifying individual agents and allowing the aggregate market behaviour to arise implicitly from their interactions at the micro-level. In most financial contexts, this amounts to simulating behaviour at the most granular level, i.e. orders submitted to the limit order book (LOB), which is the collection of unmatched orders submitted by market participants. However, as such models are mostly based on pre-defined agent policies, previous approaches have been criticised for having too simplistic agent strategies and emergent behaviour. Reinforcement Learning (RL) has been proposed as a potential solution \cite{frey2023jax,ardon2021towardsRLbased,Spooner-Market-Making-RL-2018}, with most recent work focusing on single-agent RL. To build realistic and robust ABMs with \textit{intelligent} agents -- which learn from each other --  we require multi-agent reinforcement learning (MARL) which aims to learn independent policies for a heterogeneous agent population. A natural first step is to successively train RL agents in the single-agent setting, progressively adding their learned policies to the pool of opponent strategies, a form of self-play with differing policy objectives \citep{mascioli_financial_2024}.

To allow for learning in an adversarial context, classical MARL has multiple agents independently collecting experiences and simultaneously updating their policies. To the best of our knowledge, there is no open-source framework for this in an HFT context. One reason for this is the computational restrictions and training instabilities common to MARL experiments, exacerbated by the very low signal-to-noise ratio in financial applications. To overcome these challenges, we require a highly efficient environment to enable large-scale MARL experiments. We therefore present JaxMARL-HFT, the first GPU-enabled MARL environment for HFT on the LOB. Leveraging the power of JAX \citep{jax2018github} and extending the parallelisation of JAX-LOB \citep{frey2023jax} to the multi-agent setting, JaxMARL-HFT is able to achieve up to a 240x reduction in end-to-end training times compared to state-of-the-art implementations on the same hardware. This speed-up allows for extensive hyperparameter sweeps and the processing of tens of thousands of parallel environments, thus unlocking the potential of rich market-by-order (MBO) data sets, such as LOBSTER \citep{huang2011lobster}. The environment is based on the JaxMARL framework \citep{flair2024jaxmarl}, to allow plug-and-play interactions with state-of-the-art MARL algorithms. 
It opens up HFT at the most granular level as a complex testbed for evaluating the limits of MARL algorithms.

In Section \ref{ch:background} we introduce the core concepts used in this paper, and in Section \ref{ch:related-work} we present an overview of the existing frameworks. Section \ref{ch:Design} presents the design of our implementation, and an overview of the implemented agent archetypes (Section \ref{ch:agent-types}) and the changes made to the JaxMARL reference implementation of IPPO (Section \ref{ch:ippo}). Results for the increased throughput compared to state-of-the-art implementations are discussed in Section \ref{ch:speed-benchmark}. Section \ref{ch:MARL-results} reports training results and evaluations against baselines of a simple two-player setup. In Section \ref{ch:conclusion}, we conclude with a summary of our contributions, and future work.

Our main contributions are summarised as follows:

\begin{enumerate}
\item Extension of the JAX-LOB \citep{frey2023jax} environment to the multi-agent setting, allowing for \textbf{heterogeneous agents}.
    
\item Compatibility with, and extension of, \textbf{JaxMARL} \citep{flair2024jaxmarl} algorithm implementations.
    
\item \textbf{Major throughput increase} and memory optimisation allowing for training with multiple years of LOBSTER \citep{huang2011lobster} market-by-order data.
    
\item  The code for the JaxMARL-HFT framework is available on \href{https://github.com/vmohl/JaxMARL-HFT}{GitHub}\footnote{\url{https://github.com/vmohl/JaxMARL-HFT}}.
    
\item  A challenging environment with \textbf{real-world} applications to test the limits of MARL algorithm development.
    
\item  Preliminary results showing that, under certain conditions, policies can be learned to outperform baselines, though they require carefully crafted action spaces and reward functions.
\end{enumerate}


\section{Background}
\label{ch:background}

\subsection{High-Frequency Trading}
\textbf{Limit order books (LOBs)} \citep{gould2013limit,bouchaud2018trades} are the primary mechanism through which modern financial exchanges operate. Market participants can submit limit orders to express their intention to buy or sell a particular asset at a given price and quantity. As continuous double auction markets, LOBs collect buy and sell limit orders from market participants based on price and time priority, and match compatible orders. Populations of market participants sharing objectives are often classified by trading task. Despite significant intra-class heterogeneity, the formulations of \textit{market making, order execution} and \textit{directional trading} prove useful in understanding general population characteristics \citep{gould2013limit,bouchaud2018trades}. These categories provide mathematical frameworks from which control solutions have been designed.


\textit{Market making} refers to participants who provide liquidity to an exchange by quoting bid and ask prices, aiming to profit from the spread whilst minimising risks associated with accumulating an inventory of securities.

\textit{Order execution} is a task performed by participants looking to buy or sell a specified quantity of a security over a pre-determined period. The participants in this task aim to minimise the costs associated with this transaction.

\textit{Directional trading} refers to the general case where a participant aims to profit from short-term price movements by buying and selling a security without the explicit aim for liquidity provision as in the market making task.

\subsection{Multi-Agent Reinforcement Learning (MARL)}
Multi-agent reinforcement learning is reinforcement learning where multiple agents learn to act in a common environment simultaneously \citep{flair2024jaxmarl}. The increased computational requirements for training multiple agents simultaneously have typically limited MARL developments. Libraries such as JAX \citep{jax2018github} provide a Python interface through which researchers can easily implement hardware acceleration on the GPU and just-in-time (JIT) compilation. Developing MARL environments compatible with JAX acceleration is an active research domain, whilst traditionally environments were designed for the CPU. \citet{flair2024jaxmarl} present the JaxMARL library to address this problem; they provide a range of open-source MARL environments leveraging JAX for GPU acceleration.

\section{Related Work}
\label{ch:related-work}

Early research in the domain of agent-based modelling for financial markets demonstrates that interactions between basic zero-intelligence agents are sufficient to replicate some key market phenomena \citep{palmer1994artificial, Gode-1993-Allocative} and are capable of explaining a substantial share of the cross-sectional variation in bid-ask spreads \citep{farmer2005predictive}. However, later studies indicate that capturing certain more nuanced market dynamics, such as price impact \citep{cui-price-impact} or order flow correlation \citep{Vyetrenko-get-real}, may require agents with more intelligent decision-making capabilities.

More recently, (deep) reinforcement learning has been increasingly applied to different financial trading problems, including in particular directional trading \citep{theate2021application, liu2020adaptive}, market making \citep{Spooner-Market-Making-RL-2018, gavsperov2021reinforcement} and order execution \citep{Moallemi03062022, ning2021double}. Some studies have already taken the first steps towards utilising the powerful learning capabilities of RL agents as a more sophisticated version of agent-based modelling for financial markets by leveraging MARL. \citet{lussange_modelling_2021} show that a trading simulation with multiple RL agents, based on daily stock-market data, can reproduce several market statistics, specifically price returns, volatility at different horizons, and autocorrelation metrics. In a subsequent study, they support these findings by extending their analysis to daily cryptocurrency closing price data \citep{lussange2024modelling}. \citet{ardon2021towardsRLbased} train multiple RL agents that act either as liquidity takers or as liquidity providers, coupled with a stochastic background model following \citet{cont2021stochastic}, and demonstrate that groups of agents can learn a wide spectrum of different behaviours. Extending this approach, \citet{yao2024reinforcement} also employ agents of those two types, but remove the background order‑book model. Using one day of tick‑level order book data from the LOBSTER dataset \citep{huang2011lobster}, they show that the resulting simulation reproduces several stylised facts observed in the real data. \citet{yao2024reinforcement} also highlight the extensive computational resources required for their training setup. These computational requirements have generally limited the scale of previous MARL research for HFT, or enforced simplified approaches such as stochastic background models instead of data-driven development.

High computational cost and long training times are common problems in MARL applications beyond finance, motivating the development of JaxMARL \citep{flair2024jaxmarl}, a JAX‑based \citep{jax2018github} framework that provides a range of classical MARL algorithms and environments. Our work extends JaxMARL and unlocks the benefits of GPU-accelerated MARL for a real‑world–relevant and highly challenging application: modelling financial markets at HFT scales. Based on the JAX-LOB simulator \citep{frey2023jax}, we present a flexible MARL environment that delivers a substantial RL training speed‑up. 
Using GPU acceleration to advance MARL research has been used in application domains beyond the environments presented in JaxMARL, most notably for self-driving, where it has seen significant success \citep{cusumano-towner2025robust,kazemkhani2025gpudrive}. Nevertheless, these examples were not implemented in JAX, but based on an engine \citep{shacklett23madrona} written in C++ code compiled to the GPU.

We open-source our implementation, which places it within the current lineage of publicly available ABM frameworks for high-fidelity LOB trading, such as MAXE \citep{FastAgent22Belcak}, ABIDES \citep{byrd_abides_2020}, and PyMarketSim \citep{mascioli_financial_2024}. These frameworks support heterogeneous agents that submit synthetic orders to a simulated limit‑order book, enabling controlled studies of market microstructure. ABIDES is complemented by the ABIDES‑gym extension \citep{abides-gym}, which adds an interface for deep reinforcement learning and has already underpinned RL studies on optimal order execution \citep{karpe_multi-agent_2020,lin_agent-based_2021} and market making \citep{vicente2022DeepQ}. PyMarketSim \citep{mascioli_financial_2024} also provides a limit‑order‑market environment populated by a range of heuristic traders and additionally offers training strategic agents with deep reinforcement learning. It also supplies a simplified MARL facility via the Policy‑Space Response Oracles procedure, whereby, in each iteration, a single best‑response policy is trained via reinforcement learning against the current mixed‑strategy equilibrium, then frozen and added to the global strategy pool before the equilibrium is recomputed. Overall, JaxMARL-HFT departs from existing ABM frameworks in two key aspects: it leverages GPU-acceleration for a significant RL training speed-up, and adopts a classical, fully concurrent MARL paradigm, enabling researchers to use existing algorithm implementations, such as those available in JaxMARL.

\section{Design}
\label{ch:Design}

We design JaxMARL-HFT as a LOB-level MARL environment in JAX \citep{jax2018github}. It is built on JAX-LOB \citep{frey2023jax} and it is compatible with JaxMARL \citep{flair2024jaxmarl}. It follows a classical MARL framework, enabling multiple reinforcement-learning agents to interact both with each other and with historical LOB message streams.

Leveraging JAX has three key performance advantages \citep{jax2018github,flair2024jaxmarl}. First, JAX's vectorised mapping (\texttt{vmap}) allows for parallelisation across thousands of GPU threads. Second, the framework benefits from JAX's just-in-time (JIT) compilation, which automatically fuses operations into an optimised kernel and eliminates Python overhead. Third, as both environment rollouts and learning updates are executed on the GPU, the data-transfer latency arising between CPU and GPU in conventional implementations is removed.

 \begin{figure}[htbp]
  \centering
  \includegraphics[width=0.9\linewidth]{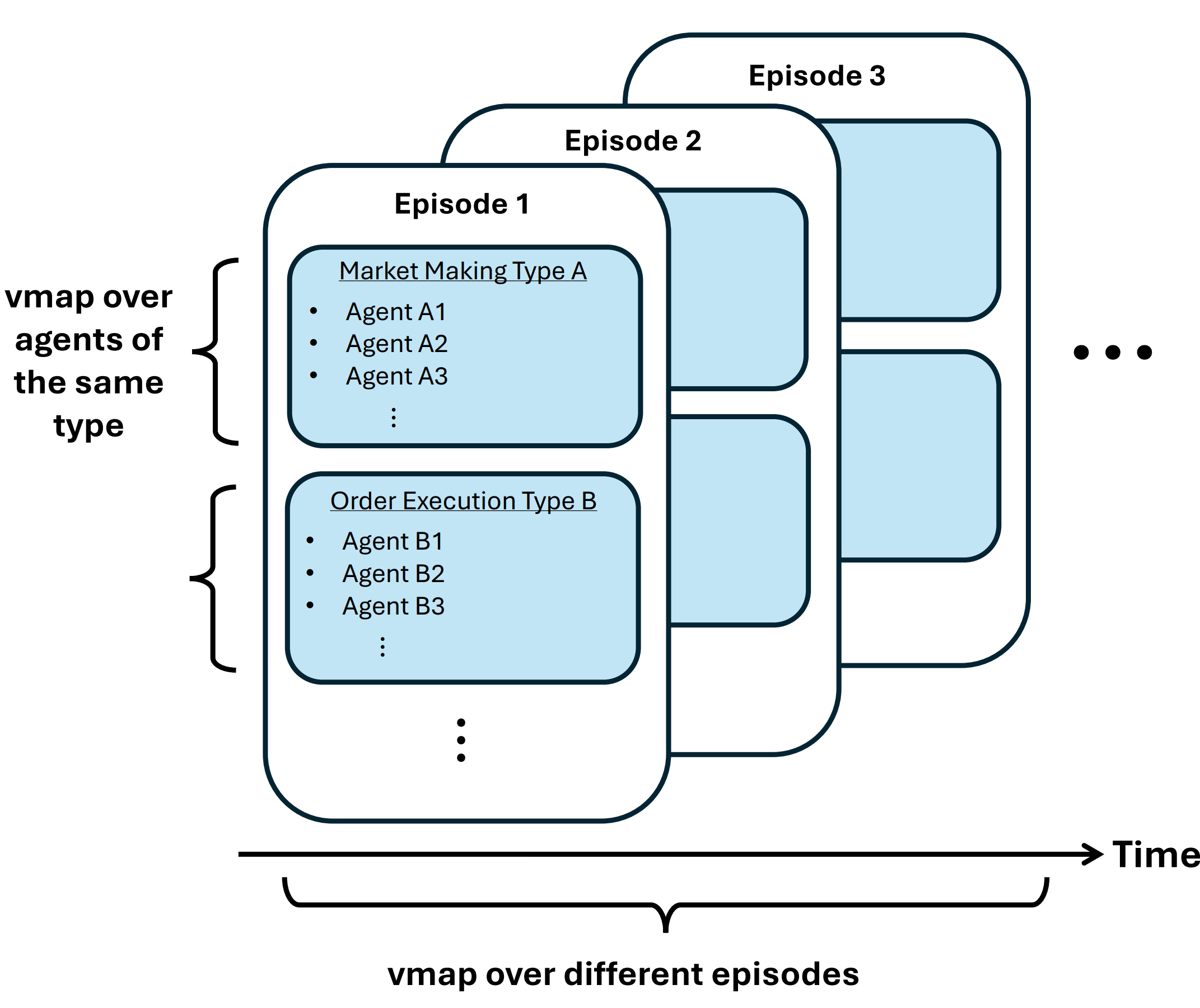}
  \caption{The two levels of parallelisation on the GPU with vectorised mapping (\texttt{vmap}) in JAX.}
  \label{fig:vmap_design}
\end{figure}

We employ the \texttt{vmap} parallelisation in two different ways, as illustrated in Figure \ref{fig:vmap_design}. First, we partition the dataset into individual episodes. The start and the length of each episode can be chosen flexibly; episodes may even be specified to overlap. At the beginning of each episode, the LOB is initialised as described in \citet{frey2023jax}, using the current state of the book at that time. With \texttt{vmap}, we can then process these episodes in parallel. We make a significant improvement compared to JAX-LOB in how messages are loaded on the GPU memory at initialisation. Rather than loading the data in windows, as described in Figure 1 of \citet{frey2023jax}, we load all messages of the dataset in contiguous form and instead keep careful track of the indices where episodes start, and index the data at runtime. This avoids excessive padding and allows a year's worth of pre-processed AMZN order data to occupy just 4GB of GPU memory, which was not possible in the original implementation.

Second, we parallelise across agents of the same type, at every step within an episode. The reason we do not \texttt{vmap} over all agents indiscriminately is JAX-specific: the underlying arrays of the functions (e.g. generating observations or constructing the LOB message based on the action of an agent) must have identical shapes across agents. Since JaxMARL-HFT is designed with the goal of maximum flexibility, allowing for heterogeneous observation spaces, action spaces and reward functions, this is generally not the case. One possible workaround would be the utilisation of padding, but we aim to keep the possibility for highly diverse agents, e.g. those whose observation spaces only contain a handful of hand-coded features versus those whose observations comprise tokenized messages that could be orders of magnitude larger. In these cases, padding would be extremely inefficient. To preserve flexibility, we iterate in an outer Python for-loop over agent types and \texttt{vmap} only over instances of the same type, as depicted in Figure \ref{fig:vmap_design}. This procedure strikes a good balance: it makes use of the benefits of GPU parallelisation, while maintaining support for heterogeneous agents.


A single environment step is structured as follows: 
\begin{enumerate}
\item \textbf{Action Conversion:}
        Agent actions are transformed into LOB messages based on the chosen action space. Unless new messages are at the same price-level, cancellation messages are prepared for unmatched orders from the previous step. 

\item  \textbf{Random shuffling:}
        All agent messages are randomly shuffled, so there is no ordering in how agents act.
        
\item  \textbf{Augment with market-replay messages:}
        The cancellation and action messages generated by the agents are concatenated with a background stream of historical messages, in this order. Because the environment is modular, these can be replaced by another background model, e.g. a generative model such as in \citet{NagyGenerativeAI2023}, with some additional implementation.
        
\item \textbf{Processing by JAX-LOB:} Orders are processed by the JAX-LOB simulator as described in \cite{frey2023jax}, and the resulting trades are recorded.
 
\item  \textbf{Calculate step outcomes:}
        Rewards, auxiliary information, termination flags, and  observations are computed. These are used by the agents to determine their next action.
\end{enumerate}

\subsection{Different Agent Types}
\label{ch:agent-types}

Two heterogeneous agent categories are implemented, covering  three HFT tasks, each comprising a set of action spaces, observation spaces, and reward functions.

\subsubsection{Market Making}\label{sec:mm-impl} The market making agent implementation supports three action space configurations, providing a diverse set of strategies and a range of task complexities. The \textit{Spread-Skew} action space is a simplified version of the action space used by \citet{ardon2021towardsRLbased}, with tabular actions used in place of spread-skew parameters. The \textit{Fixed Quantity} action space is inspired by the work of \citet{Spooner-Market-Making-RL-2018}. Finally, the \textit{AvSt} action space is a parametrised form of the optimal control solution presented by \citet{avellaneda2008high}, similar to the implementation in \citet{ardon2021towardsRLbased}.

In these action spaces, the agent selects bid and ask prices to present to the market from a pre-determined selection of strategies. For example in the \textit{Fixed Quantity} space, the agent has 8 options; (i) Not trading; (ii)/(iii) Posting 2 and 4, respectively, ticks away from the best prices; (iv) trading one tick into the spread; (v)/(vi) Posting 2 ticks away from the spread on either side of the book;  (vii)/(viii) Posting 5 ticks away from the best price on one side, and 1 tick into the spread on the other.

The quantity associated with each order is fixed a priori through an environment configuration variable: the agents following these strategies will quote constant volumes associated with each bid and ask message. In practice, this quantity can be set to the minimum lot size, which, depending on the asset price, is often the most commonly used trade size by market participants, making this assumption reasonably realistic. 

Similar to our approach to action spaces, we design flexible reward functions. We include rewards based on \citet{vadori_towards_2024} and \citet{Spooner-Market-Making-RL-2018}, which include \textit{spread} and \textit{inventory} PnL terms, with variable weights given to each term, to allow diverse agent risk tolerances. All functions are parameterised by adjustable hyperparameters, which presents the user with a highly diverse set of possible reward functions.

For the observation space, the market making agents are presented with a one-dimensional array containing statistics related to the agents' task, such as the size of their current inventory. We implement multiple similar observation spaces, with a range of complexities, allowing the user to select an observation space with a level of detail appropriate for their use case.

\subsubsection{Order Execution}
\label{ch:order-exec-agent}
We base the order execution agent on the environment presented by \citet{frey2023jax}. We discretise the action space, so that rather than selecting a quantity to post at each of the four reference prices, the agent selects only at which price to submit an order of a quantity defined a priori.  A more complex version extends this, to additionally allow orders with multiples of 2 or 5 of this quantity. The pre-defined order quantity is selected based on the execution task size and episode length.

\subsubsection{Directional Trading}
Finally, we implement a directional trading action space as a configuration of the market making class, re-using all reward functions and observation spaces. The similarities between these tasks justify this approach; however, we develop a unique action space, to ensure diversity between the classes. The directional trading action space enables the agent to either send a bid or ask order at the best price, or do nothing at each step. All bid and ask messages quote a fixed quantity.

\subsection{IPPO and Heterogeneous Agent Types}
\label{ch:ippo}
JaxMARL \citep{flair2024jaxmarl} includes MARL algorithm implementations. Given the different classes of agents described previously, we need to extend these implementations to be compatible with our environment. This is relatively easy for an algorithm like independent PPO, as we simply maintain separate networks, observations, hidden states, and actions for each agent type, iterating over them during rollout and at update steps. Agents of the same type continue to benefit from efficient batch operations during training, as they do for environment rollouts. We opt not to go into great detail on the algorithmic details of IPPO, but 
highlight that all agents retain separate network weights, and thus only learn about adversarial behaviour from environment observations, not through shared weights.

\section{Experimental Setup and Results}

\begin{table*}[!hbtp]
  \caption{Speed comparison between JaxMARL‑HFT, PyMarketSim, ABIDES, and CPU‑MARL. For JaxMARL-HFT, ABIDES-gym and CPU-MARL we used the RL pipeline with a random policy. We parallelise across 4000 environments (50 steps per environment) on a single GPU for JaxMARL-HFT and 64 CPU cores (one thread per core) for ABIDES-gym and CPU-MARL. For PyMarketSim, we use their non-parallelised, purely zero-intelligence (ZI) agents-based simulation on 1 CPU core, matching the number of ZI agents to the sum of messages and random-policy agents in the other settings. JaxMARL-HFT and the CPU-MARL employ 2 types of agents (with a varying number of agents), whereas for PyMarketSim and ABIDES there is only 1 agent. Since PyMarketSim and ABIDES-gym do not offer classical MARL, the corresponding entries for multiple agents per type are left blank.}

  \label{tab:speed-comparison}
  \centering
  \begin{tabular}{*{7}{c}}
    \toprule
    \multicolumn{4}{c}{\textbf{JaxMARL-HFT}}
      & \textbf{PyMarketSim}
      & \textbf{ABIDES-gym}
      & \textbf{CPU-MARL} \\
    \cmidrule(lr){1-4}\cmidrule(l){5-5}\cmidrule(l){6-6}\cmidrule(l){7-7}
     Data Messages per Step  & Agents per Type & Time (s) & Steps/s
         & \multicolumn{1}{c}{Steps/s}
         & \multicolumn{1}{c}{Steps/s}
         & \multicolumn{1}{c}{Steps/s} \\
    \midrule
    100  &  1 &  9.104 &  21969 &   463 &   734         & 1805 \\
       1 &  1 &  0.570 & 351119 & 10830 & 2979    & 4896 \\
    100  &  5 & 13.513 &  14801 &   - &   -            & 84 \\
      1  &  5 &  2.246 &  89062 &  - &   -            & 334 \\
    100  & 10 & 18.650 &  10724 &   - &   -            & 30 \\
       1 & 10 &  4.140 &  48312 &  - &   -            & 114 \\
    \bottomrule
  \end{tabular}
\end{table*}

\subsection{Speed Benchmarking}
\label{ch:speed-benchmark}
To properly evaluate our environment, we compare it to closely related state‑of‑the‑art implementations. Some of the most similar publicly available frameworks that support multi-agent simulations in HFT are ABIDES-gym \citep{abides-gym} and PyMarketSim \citep{mascioli_financial_2024}. Although these implementations permit simulations using multiple heterogeneous background agents with pre-defined policies, both focus on updating a single reinforcement learning agent; additional agents may interact with the learning agent but do not learn concurrently. Investigating the source-code of ABIDES-gym shows that future support for MARL was planned but does not seem currently supported. PyMarketSim takes a step closer to MARL through its policy‑space response oracles, which can be used to train successive RL agents, fixing their policies and adding them to a pool of possible policies, but does not allow for simultaneous learning. We 
compare to ABIDES-gym and PyMarketSim, despite the inexact match, as these represent the most similar open-source baselines. Additionally, to obtain an exact speed benchmark and to quantify the contribution of GPU acceleration, we implement a CPU version of our MARL environment, which we refer to as CPU-MARL.

We use the open-source code of ABIDES‑gym and PyMarketSim to compare all implementations on the same hardware. For our experiments, we use a fairly typical compute node for deep learning in academic contexts with up to 8 NVIDIA L40S GPUs and an AMD EPYC 9554 processor with 64 cores (256 threads) available.

\subsubsection{Speed of the Environment}

Table \ref{tab:speed-comparison} depicts the environment‑step throughput without learning updates. To ensure a fair comparison, we aim to harmonise the dynamics and configuration settings across frameworks, since they differ in several implementation details. In JaxMARL‑HFT, in each step both historical market‑replay LOBSTER data messages and messages generated by RL agents are processed. For JaxMARL-HFT and CPU-MARL, we consider a different number of agents of two different types: market making and order execution. In the results shown in Table \ref{tab:speed-comparison}, these RL agents choose random actions and no learning updates are performed. PyMarketSim, by contrast, does not use market‑replay data. Instead, market activity is generated by background agents (e.g., zero‑intelligence agents) that submit orders. We therefore instantiate PyMarketSim with the same total number of zero‑intelligence agents as the sum of data messages and random RL agents in JaxMARL‑HFT and CPU-MARL. Because agent arrival times in PyMarketSim follow a geometric distribution, we adjust this distribution so that, in expectation, every agent acts in every step. The simulation using only background agents is not parallelised across CPU cores in PyMarketSim, however their reinforcement learning process - which is based on the \textit{Tianshou} package of \citet{weng2022tianshou} - is parallelised, and we report a comparison with their parallelised RL pipeline across 64 cores in Section \ref{ch:RL-Speed}. For ABIDES-gym, environment step frequency is not defined by number of messages, but by simulation time, we use the rmsc04 reference configuration and empirically adjust the inter-step time to have 1 and 100 orders processed on average. As illustrated in Table \ref{tab:speed-comparison}, JaxMARL-HFT achieves a substantial speed-up relative to current state-of-the-art reference implementations and the comparable CPU-based MARL environment, which seems to increase with the number of agents.

\subsubsection{Speed of Multi-Agent Reinforcement Learning Training}
\label{ch:RL-Speed}

\begin{figure}[!h]
  \centering
  \includegraphics[width=0.95\linewidth]{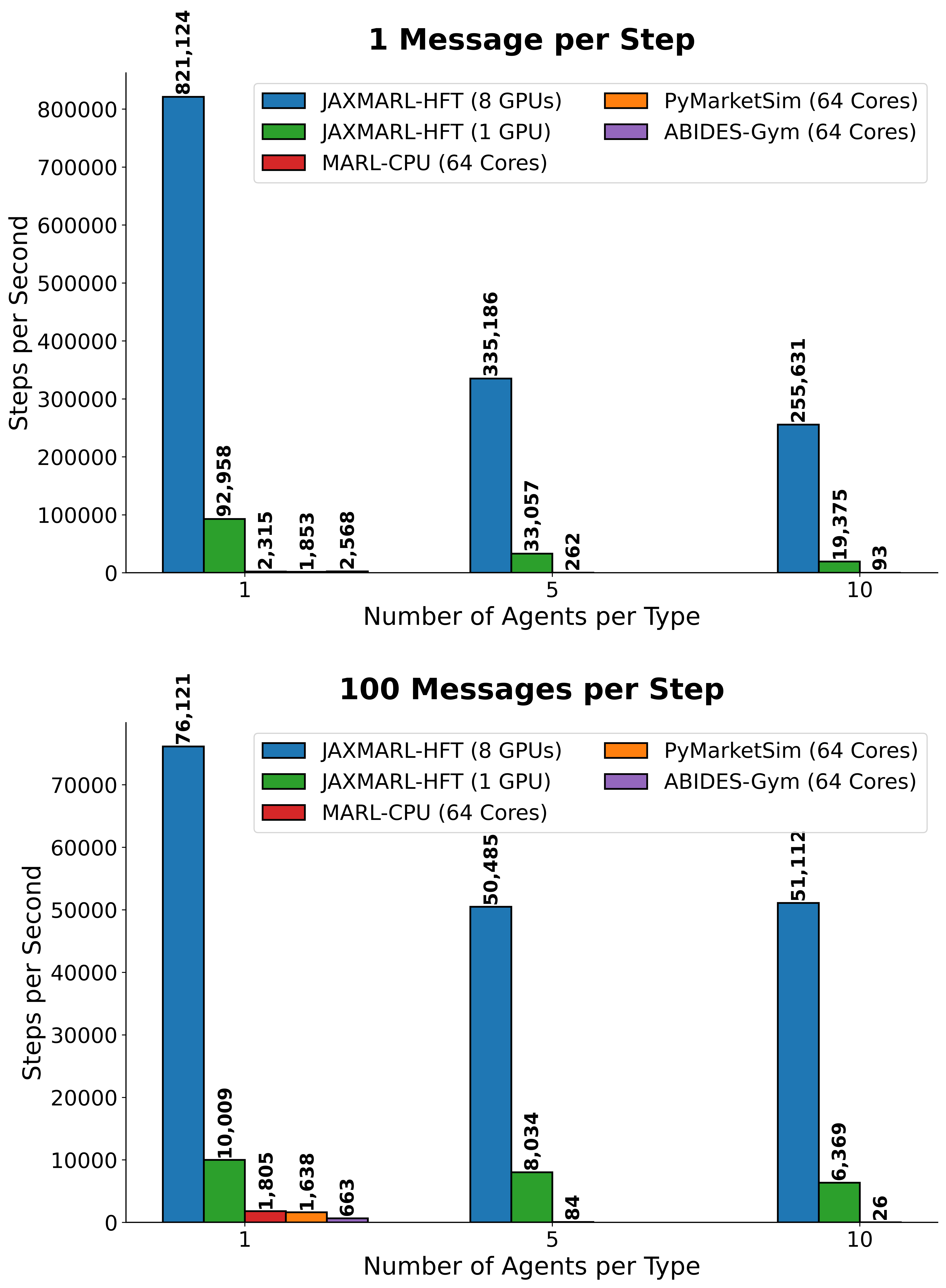}

  \caption{Speed Comparison of the Reinforcement Learning Training Pipeline}
  \Description{Two bar plots comparing steps per second for different frameworks. In the first plot, one background message is replayed per step, and in the second plot one hundred background messages per step. Across both plots, JaxMARL-HFT achieves much higher throughput than comparable CPU-based implementations. The difference grows when moving from one hundred to one background message per step and when increasing the number of agents. }
  \label{fig:RL-Speed}
\end{figure}

The key advantages of GPU-acceleration become apparent when considering the MARL training loop. Unlike current state-of-the-art frameworks, which typically execute environment rollouts on the CPU and perform policy updates on the GPU, our system carries out both the environment rollouts and the RL updates entirely on the GPU. 

With respect to the results in Figure \ref{fig:RL-Speed}, it is important to emphasise that, in this comparison, only JaxMARL-HFT and CPU-MARL conduct MARL with several learning agents concurrently. By contrast, the results for PyMarketSim and ABIDES-gym each encompass only a single agent updating its policy. We endeavoured to make the comparison as fair as possible and to harmonise hyperparameters. All implementations perform their RL updates on an NVIDIA L40S GPU. ABIDES-gym, PyMarketSim, CPU-MARL additionally use an AMD EPYC 9554 processor for their environment rollouts (64 parallel environments on 64 cores with 1 thread per core, i.e. no simultaneous multithreading). For JaxMARL-HFT 4096 environments are used in parallel on the GPU.

Figure \ref{fig:RL-Speed} depicts the significant speed advantages of a fully GPU-accelerated framework. The performance benefit becomes particularly pronounced as the number of learning agents increases. While the speed-up for a single agent of each type is approximately 5x - 35x, it rises to 95x - 125x for five agents. For ten agents, we even observe a 200x - 240x speed improvement. Additionally, a JAX-based implementation makes it straightforward to distribute across multiple GPUs through leveraging JAX's \texttt{pmap} capabilities. Comparing the implementations when the entire node is used, i.e. 8 GPUs and 256 workers benefiting from multi-threading, the throughput increases by a factor of 50 when comparing JaxMARL-HFT to ABIDES-gym in the single-agent case for 100 messages per step.\\

We notice significant differences in the increased throughput between implementations when varying the number of LOB messages processed per RL-step. For example, comparing 1 GPU to the next-best PyMarketSim, we see an increase in throughput of 5.5x versus 40x for 100 and 1 messages, respectively. This strongly underlines the obtained benefits when frequent model inference steps are required on the GPU: data does not need to be continuously transferred between devices in the JaxMARL-HFT implementation.

\subsection{MARL: Independent PPO for High-Frequency Trading}\label{ch:MARL-results}

\subsubsection{Training curves and learned policies}
Using the adapted IPPO framework (Section \ref{ch:ippo}), a two-player environment is considered, containing a market making and an execution agent, learning simultaneously in the MARL environment. An overview of the specifications for each of the agents is given in Table \ref{tab:parameters}. We use a GRU-based network architecture to ensure recurrence can identify time-dependent patterns in the data, as described in \citep{frey2023jax}.

\begin{table}[h!]
\caption{Key parameters used by the agents considered in Section \ref{ch:MARL-results}, where square brackets indicate multiple options were considered in the presented results.}
\begin{tabularx}{\linewidth}{>{\bfseries}l X}
\toprule
Parameter & Value \\
\midrule
Dataset & AMZN 2024\\
Episode start frequency & Every 64 steps\\
Episode length & 64 steps\\
Data messages/step & 100\\
JAX-LOB book capacity & 100 orders\\
MM: Action Spaces & [SpreadSkew, FixedQuant]\\
MM: Reward Functions & [BuySell, Spooner]\\
MM: Value reference price & [Mid Price, Best Price]\\
MM: Inventory penalty  & [Quadratic, None]\\
MM: Quadratic penalty fact. $\rho$ & 50\\
MM/EXEC: Order size & 10 \\
EXEC: Task direction & Uniformly sampled\\
EXEC: Action Space & Complex (Sec. \ref{ch:order-exec-agent})\\
EXEC: Reward func. & See \citet{frey2023jax}\\
EXEC: Reward $\lambda$ & 0.0\\
EXEC: Task size & 600\\
EXEC: Unfilled order penalty & [0.1,0.05,0.01]\\
\bottomrule
\end{tabularx}
\label{tab:parameters}
\end{table}

The nature and relatively large number of parameters in Table \ref{tab:parameters} illustrates one of the open problems with MARL for high-frequency financial applications: stable training is difficult with general-purpose action spaces and multidimensional observation spaces depicting the total state of the LOB. We therefore content ourselves, for the purpose of illustrating the functionality of JaxMARL-HFT, with strongly-simplified action and observation spaces. Multi-modality on the optimal actions is a reason for which the action spaces are discretised into a relatively small number of distinct classes, rather than allowing more general, continuous action space described in \citet{frey2023jax}. \\

\begin{figure}[h!]
    \centering
    \begin{subfigure}[t]{\linewidth}
        \centering
        \includegraphics[width=0.8\linewidth]{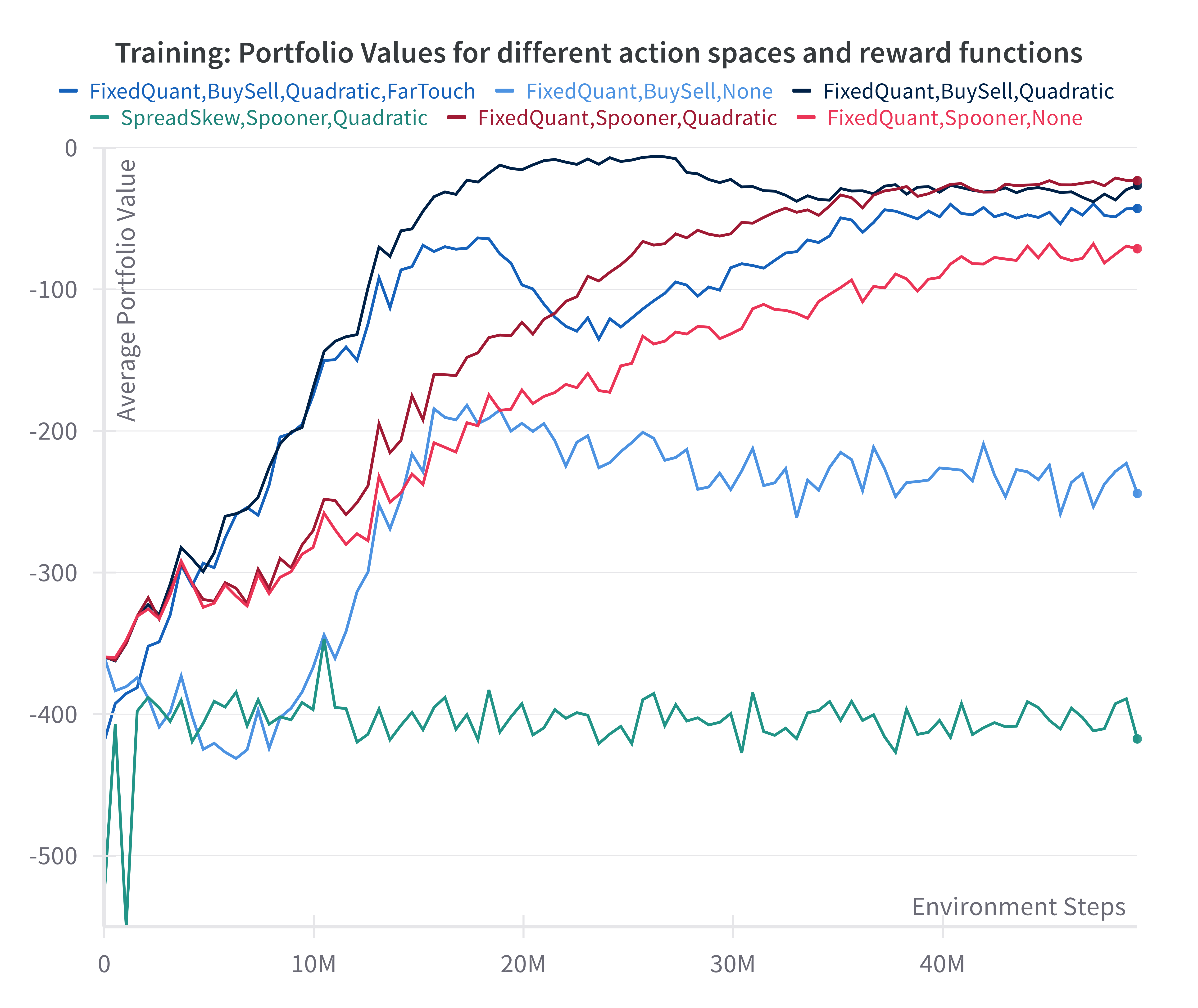}

        \caption{Market making training curve showing portfolio value. Curves are shown for different reward functions and action spaces (Red/Green/Blue). Further, the effect of a quadratic inventory penalty term and the reference price for portfolio value calculation is considered.}
        \label{fig:MM-training-curves}
    \end{subfigure}
    \begin{subfigure}[t]{\linewidth}
        \centering
        \includegraphics[width=0.8\linewidth]{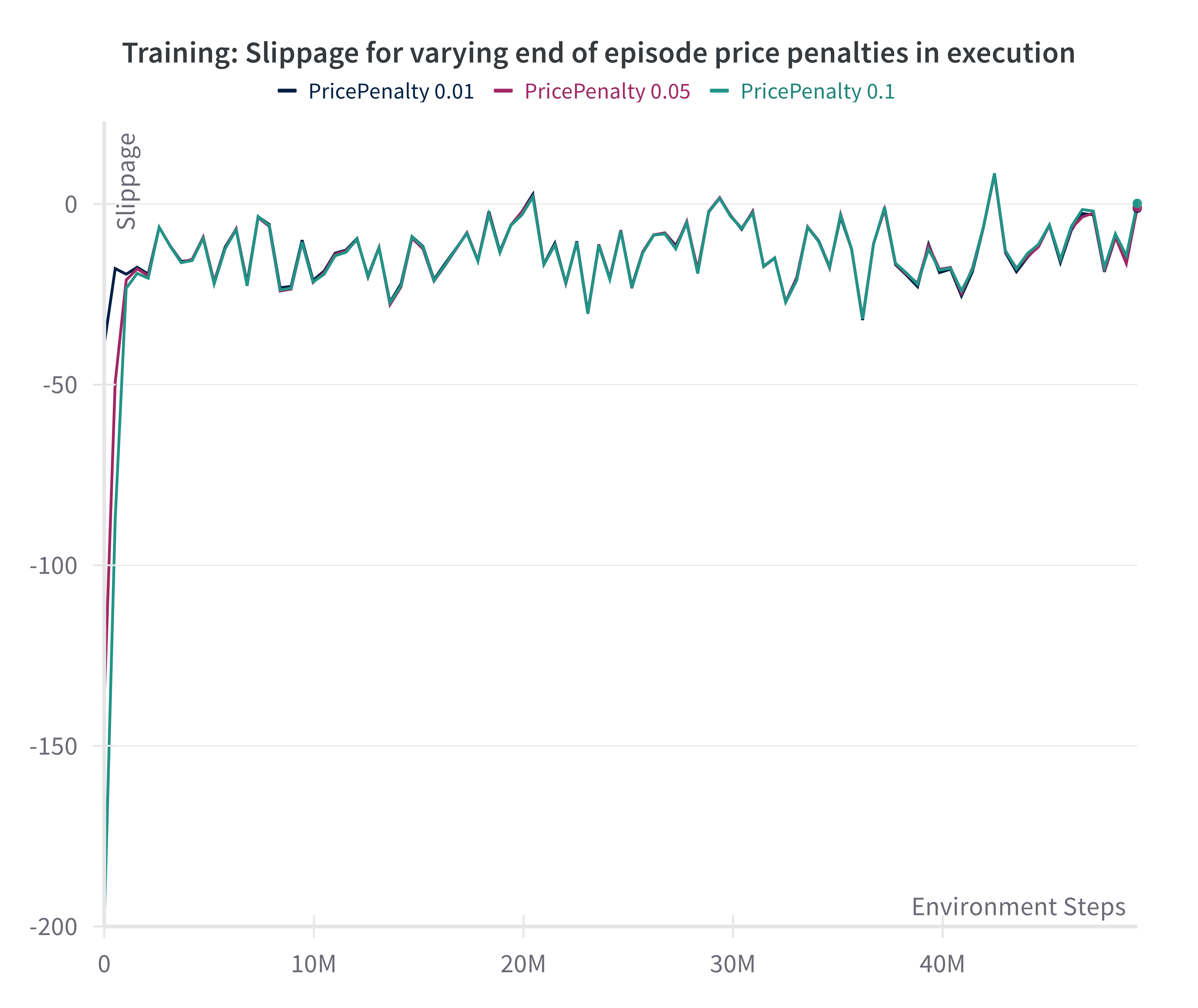}

        \caption{Execution agent training curve showing ablations over the end-of-episode reward penalty. Except for the first updates, there is no major difference as avoiding this penalty is learned very quickly.}\label{fig:EXEC-training-curves}
    \end{subfigure}
    \Description{Training curves for the market making and order execution agent. The first subplot shows the portfolio value for the market making agent under different settings of the reward function, action space, inventory penalty and reference price. The second subplot shows the slippage of the order execution agent for different end-of-episode reward penalties.}

    \caption{Training curves using a proxy reward to aid comparison across reward functions. MARL experiment with a market making and an execution agent. }\label{fig:ippo-training-curves}
\end{figure}

Nevertheless, with such settings, we observe some promising results. Figure \ref{fig:MM-training-curves} shows the training curves for two of the more promising action-spaces (SpreadSkew and FixedQuant), for market making. It is important to note that in both cases, we show a proxy value to measure the ability of the agent to execute the given task, which is not necessarily the reward function. We use the \textit{portfolio value} ($PV$) \eqref{eq:portfolio_value} as a measure for the market maker, and the \textit{Slippage} \eqref{eq:slippage} adjusted for the direction of the execution task. This is because it allows for a comparison across different reward functions. $P_{ref}$ is generally chosen to be the mid price, but we show an example in Figure \ref{fig:MM-training-curves} where the reference price is the far-touch price instead.

\begin{equation}
\label{eq:portfolio_value}
    PV = Q_{inv} \times P_{ref} + Cash
\end{equation}

\begin{equation}
\label{eq:slippage}
    Slippage = (-1)^{dir}\sum_j Q_j (P_j-P_{init})
\end{equation}

For example, in Figure \ref{fig:MM-training-curves} we show the differences in training when applying different reward functions: the 'Spooner' reward \eqref{eq:SpoonerReward} from \citet{spooner_robust_2021} (in Red), and the 'BuySell' reward based on instantaneous differences of trade prices to the mid price \eqref{eq:BuySellReward} (in Blue). Following \citet{spooner_robust_2021}, we define trading PnL terms as $\Psi_b = \sum (\bar{M} - P_b) Q_b$ and $\Psi_s = \sum (P_a - \bar{M}) Q_a $, with $\bar{M}$ representing the average mid price over the previous step and $Q$ and $P$ representing bid or ask prices and quantities, respectively. We define the inventory PnL similarly as $\Psi_{\text{INV}} = I_t (M_t - M_{t-1})$, where the configurable hyperparameter $\lambda$ controls the relative weighting of positive and negative PnL, and $M_t$ and $I_t$ denote the mid price and inventory at time step $t$. 

\begin{equation}
    \label{eq:SpoonerReward}
   r_\text{Sp}= \Psi_b + \Psi_s + \Psi_{\text{INV}} - (1-\lambda)\max(0, \Psi_{\text{INV}})
\end{equation}

\begin{equation}
    \label{eq:BuySellReward}
    r_{buy-sell}=\Psi_b +\Psi_s 
\end{equation}

Further, the effect of a quadratic reward penalty on the held inventory (Dark Blue/Red) is considered. We observe that such a penalty is beneficial, as is the use of the Spooner reward function, which itself depends on the held inventory. It is important to highlight that despite convergence of the loss and reward in both training and validation data, none of the learned strategies for market making are able to make net profits, but instead lose approximately 0.2 ticks on average. Section \ref{ch:baseline-eval} shows that this is still a surprisingly good result. It is assumed that one of the hindering factors which makes profitable market making difficult is the fact that posted orders from previous steps are cancelled at each RL step. This is a marked disadvantage in a price-time priority LOB. Addressing this requires revisiting the environment design, and is left for future work.\\

Figure \ref{fig:EXEC-training-curves} shows the training evolution in the execution environment for different coefficients for the end-of-episode penalty when the given amount is not executed successfully. Clearly, a lower value implies a larger reward early on, but the network quickly learns to avoid the costly penalty, after which the difference is negligible.

\begin{figure}[h!]
  \centering
  \begin{subfigure}[t]{\linewidth}
  \centering
    \includegraphics[width=0.8\linewidth]{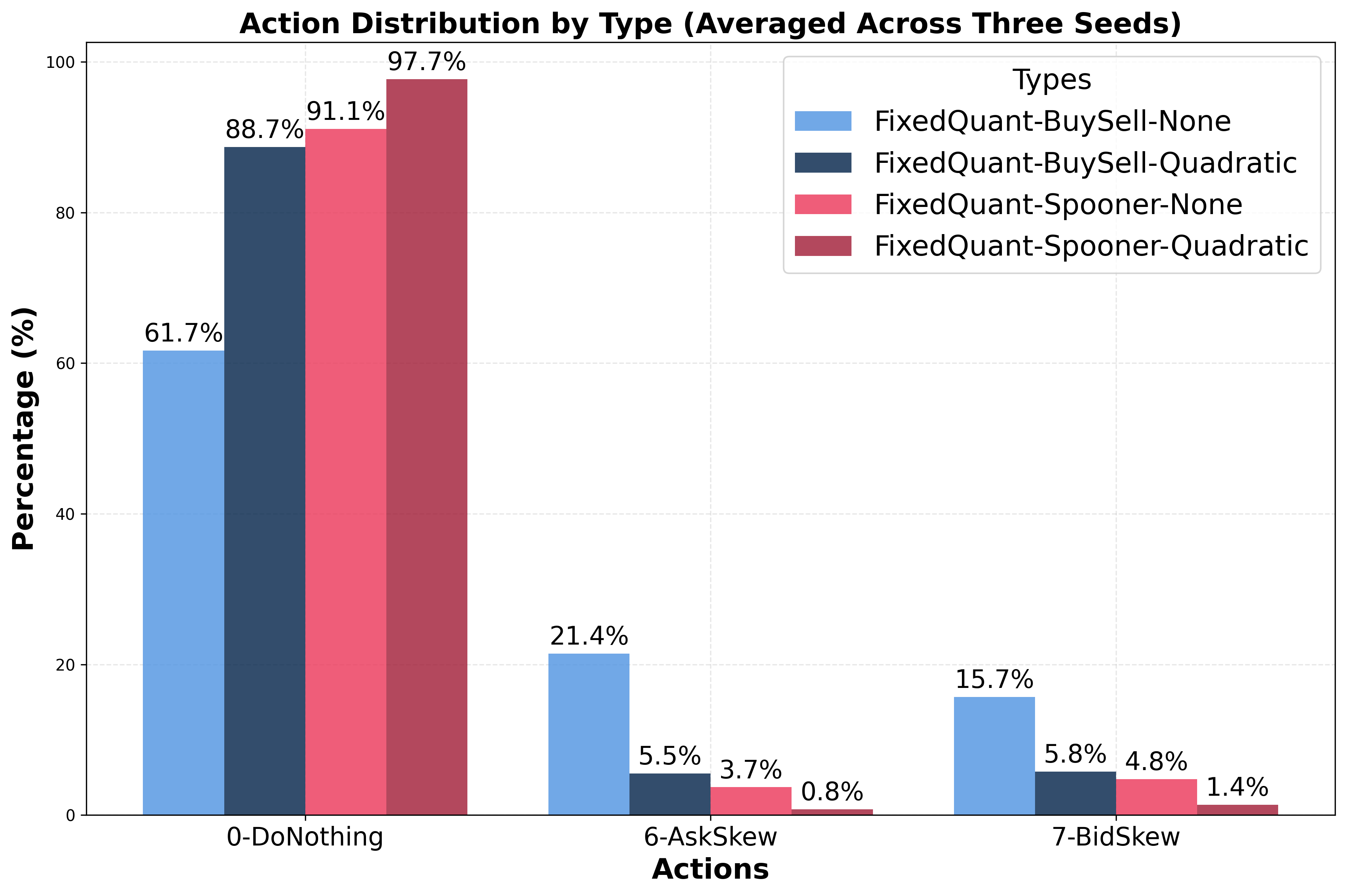}

  \caption{Market making action distribution for the three actions played with non-zero probability. Most commonly, the agent opts not to trade, this is exacerbated by an inventory penalty. The remaining time, the agent posts orders deep in the book on one side, and into the spread on the other. See Section \ref{sec:mm-impl} for details.}
 \label{fig:details-MM}
  \end{subfigure}
\begin{subfigure}[t]{\linewidth}
  \centering
    \includegraphics[width=0.8\linewidth]{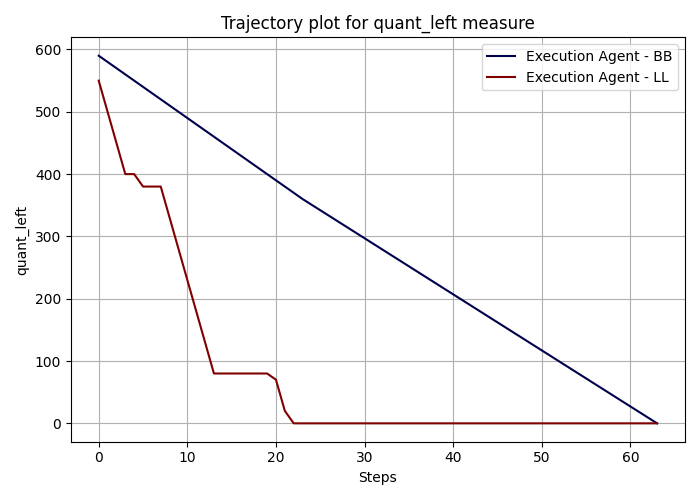}

  \caption{Example for a single episode of the learned policy behaviour compared to the baseline. The agent learns to trade much more quickly, but still has lower execution costs than an aggressive TWAP strategy, likely capturing some of the spread on occasion.}
  \label{fig:details-EXEC}
  \end{subfigure}
  \Description{Two plots showing the behaviour of the learned trading agents. The market making agent mostly refrains from trading, which is exacerbated by an inventory penalty. The remaining time, the agent posts orders deep in the book on one side, and into the spread on the other. The order execution agent trades faster than the TWAP baseline, but still has lower execution costs.}

  \caption{Detailed plots considering the behaviour of the learned policies for each agent. Results are given for policies learned using the \textit{FixedQuant} action space with the \textit{Spooner} reward function and a Quadratic inventory penalty for the market maker.}\label{fig:details}
\end{figure}

Digging deeper into the learned policies, we see that the market makers (Figure \ref{fig:details-MM}) learn to trade very infrequently, which explains the convergence towards zero portfolio value in Figure \ref{fig:ippo-training-curves}. The only actions which are sampled with non-zero frequency are strong skews with orders deep into one side of the book, and into the spread on the other. One of the drawbacks of this family of reward functions is that they are not normalised by traded volume, as in this case, an optimal policy seems to be to never trade, thus guaranteeing no losses. Choosing an action space which forces the agent to submit orders at every step is an alternative, but considering the green line in Figure \ref{fig:MM-training-curves} showing results for the \textit{SpreadSkew} action space suggests this will result in markedly reduced performance.

For execution, we see that the learned policy is generally far more aggressive than the TWAP baseline strategy (Figure \ref{fig:details-EXEC}), likely due to the end-of-episode penalty still being significant. Section \ref{ch:baseline-eval} shows that execution cost can be decreased by up to 20\% of a tick on average as compared to TWAP, likely by trading more passively, whilst still guaranteeing execution.

\subsubsection{Evaluation against baseline strategies}
\label{ch:baseline-eval}

\begin{figure}[h!]
  \centering
    \includegraphics[width=0.6\linewidth]{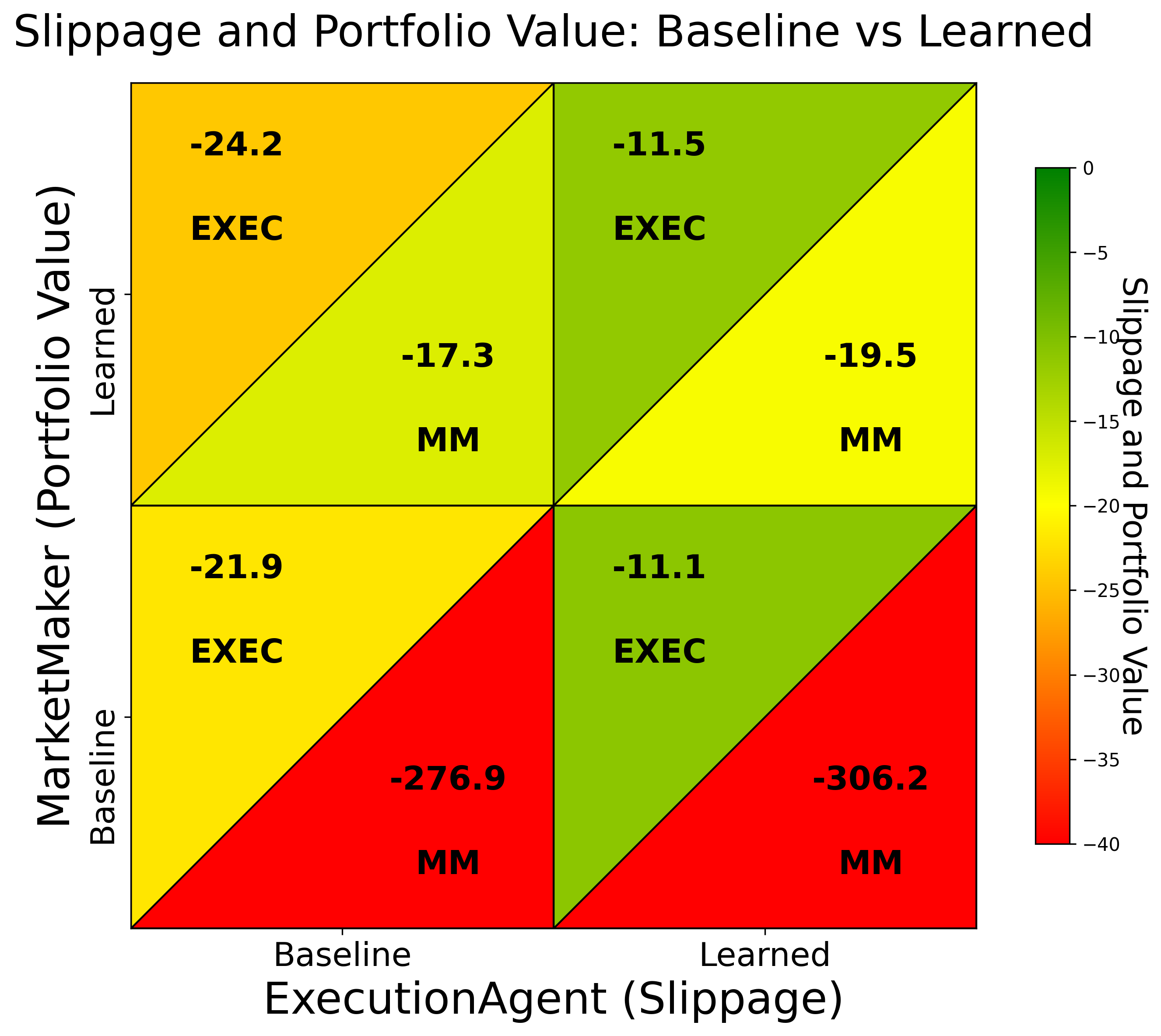}
  \Description{Matrix plot which compares the learned agents of market making and order execution with their respective baselines. Performance is shown as portfolio value for the market maker agent and slippage for the order execution agent. The learned policies outperform their baselines.}
  \caption{Evaluation of the learned agents when facing baseline implementations (TWAP and AvSt \citep{avellaneda2008high}). The bottom left has both agents play a baseline policy, whilst the top right has both play a learnt policy. The bottom triangles represent the market maker portfolio value, whilst the top represent the slippage experienced by the execution agent. The learnt policies improve over the baseline, and the execution agent performs worse when facing a learnt market making policy than when facing the baseline.}\label{fig:baseline-fig}
\end{figure}

Figure \ref{fig:baseline-fig} shows the respective improvements of the learned policies (L) using the \textit{Fixed Quant} action space with the \textit{Spooner} reward function and a quadratic inventory penalty term. The baselines (B) are TWAP in the case of the execution agent and the Avellaneda-Stoikov optimal market making model \citep{avellaneda2008high} in the case of the market making agent. Both baselines are outperformed by the learned agent in terms of their respective quality metrics (Portfolio Value and Slippage). Further, the matrix view shows that having a more strategic execution opponent decreases the performance, underlining the expected benefits of MARL when it comes to modelling indirect market impact. We leave an in-depth study of this adversarial behaviour for future work.

\section{Conclusion}
\label{ch:conclusion}

We present the first framework allowing for highly parallelised, GPU-accelerated MARL experiments for HFT. It provides a real-world-relevant and challenging multi-agent environment and is fully compatible with state-of-the-art MARL algorithms implemented in JaxMARL. The new framework contains heterogeneous implementations of the three main HFT agent tasks: order execution, market making, and directional trading. Compared to similar state-of-the-art implementations, throughput is increased by up to 240x allowing for larger models, bigger datasets, and more extensive hyperparameter sweeps. Early results show learned policies that outperform existing baseline policies and show some promising adversarial behaviour.

This novel comprehensive framework sets the stage for many future research directions: (i) Measuring the market impact obtained in simulation with learned agent policies; (ii) Moving to general-purpose action/observation spaces and only parametrising agent behaviour through the reward function; (iii) Replacing historical message data with generative models to train RL-agents in adversarial settings, even in the single-agent case; (iv) Removing historical message data after training, and considering the resulting price-series of a multi-agent model with learned agent policies; (v) As simulation allows the labelling of the origin of submitted orders, this facilitates new research in opponent shaping and market participant classification.

\begin{acks}
AC acknowledges funding from a UKRI AI World Leading Researcher Fellowship (grant EP/W002949/1). AC and JF acknowledge funding from a JPMC Faculty Research Award. We thank Peer Nagy, Bidipta Sarkar and Zilin Wang for valuable feedback and discussions.
\end{acks}
\bibliographystyle{ACM-Reference-Format}
\bibliography{references}


\end{document}